\documentclass[aps,prd,twocolumn,nofootinbib,floatfix]{revtex4} 

\usepackage{graphics}
\usepackage{graphicx}

\usepackage{latexsym}
\usepackage[cp866]{inputenc}
\usepackage[english]{babel}

\input epsf

\begin{document}

\title{\boldmath Isotopic Symmetry Breaking in the $\eta(1405)$\,$\to$\,$f_0
(980)\pi^0$\,$\to$\,$\pi^+\pi^-\pi^0$. Decay through a $K\bar K$
Loop Diagram and the Role of Anomalous Landau Thresholds}
\author{N.~N.~Achasov and G.~N.~Shestakov}
\affiliation{Laboratory of Theoretical Physics, Sobolev Institute of
Mathematics, Siberian Branch, Russian Academy of Sciences,
Novosibirsk, 630090 Russia}

\begin{abstract}
Anomalous isotopic symmetry breaking in the $\eta(1405)\to
f_0(980)\pi^0\to\pi^+\pi^-\pi^0$ decay through a mechanism featuring
anomalous Landau thresholds in the form of logarithmic triangle
singularities, i.e., through the $\eta(1405)\to (K^*\bar K+\bar
K^*K)\to(K^+K^-+K^0\bar K^0)\pi^0\to f_0(980)\pi^0 \to\pi^+\pi^-
\pi^0$ transition, has been analyzed. It has been shown that this
effect can be correctly quantified only by taking into account the
nonzero $K^*$ width. Different scales of isotopic symmetry breaking
associated with the $K^+-K^0$ mass difference are compared.
\end{abstract}

\maketitle


In the measurements of the $J/\psi\to\gamma\pi^+\pi^-\pi^0$ and
$J/\psi\to\gamma\pi^0\pi^0\pi^0$ decays carried out by the BESIII
collaboration in 2012, a resonant peak at $\sim 1.4$ GeV with a
width near 50 MeV was revealed in the three-pion mass spectra
\cite{Ab2}. Additionally, a narrow structure with a width near 10
MeV was observed in the corresponding $\pi^+\pi^-$ and $\pi^0\pi^0$
mass spectra at $\sim990$ MeV near the $K^+K^-$ and $K^0 \bar K^0$
thresholds \cite{Ab2}. Thereby, the isospin-violating decay
$J/\psi$\,$\to$\,$\gamma\eta(1405)$\,$ \to$\,$\gamma f_0(980)\pi^0$
with the subsequent $f_0(980)\to $\,$\pi^+\pi^-,\pi^0\pi^0$ decay
was observed for the first time (with a statistical significance
more than $10\sigma$). Its branching ratio was measured in
\cite{Ab2} as
\begin{eqnarray}\label{Eq5-3-1}
BR(J/\psi\to\gamma\eta (1405)\to\gamma
f_0(980)\pi^0\to\gamma\pi^+\pi^-\pi^0)=\nonumber
\\ =(1.50 \pm0.11\pm0.11)\times10^{-5}\,.\qquad\qquad\qquad
\end{eqnarray}
Taking into account the PDG data, the BESIII collaboration obtained
the ratio \cite{Ab2}
\begin{eqnarray}\label{Eq5-3-2} \frac{BR(\eta
(1405)\to f_0(980)\pi^0\to\pi^+\pi^-\pi^0)}{BR(\eta
(1405)\to a^0_0(980)\pi^0\to\eta\pi^0\pi^0)}= \nonumber \\
=(17.9\pm4.2 )\%\,,\qquad\qquad\qquad
\end{eqnarray}
which practically rules out attributing the observed isotopic
symmetry breaking to the $a^0_0(980)-f_0(980)$ mixing through the
$a^0_0(980)\to(K^+K^-+K^0\bar K^0)\to f_0(980)$ transition. At the
same time, the observation of a narrow resonance-like structure near
the $K^+K^-$ and $K^0\bar K^0$ thresholds in the $\pi^+\pi^-$ and
$\pi^0 \pi^0$ mass spectra in the $\eta(1405)$\,$\to$\,$\pi^+
\pi^-\pi^0$, $\,\pi^0\pi^0\pi^0$ decays suggests that the mechanism
of $f_0 (980)$ formation in the $\eta(1405)$\,$\to$\,$f_0(980)
\pi^0$\,$ \to$\,$3\pi$ decay is similar to that of the $a^0_0(980)
-f_0(980)$ mixing \cite{ADS79,ADS81,AS17}. In other words, this
mechanism is described by the $\eta(1405)\to(K^+ K^-+K^0\bar K^0)
\pi^0\to f_0(980)\pi^0\to3\pi$ transition, whose amplitude is
nonzero owing to the $K^+-K^0$ mass difference and is significantly
large in a narrow region between the $K^+K^-$ and $K^0\bar K^0$
thresholds.

Comparing the BESIII result (\ref{Eq5-3-1}) with the PDG data
\cite{PDG16} for the dominant decay channel $J/\psi\to\gamma
\eta(1405/1475)\to\gamma K\bar K\pi$,
\begin{eqnarray}\label{Eq5-3-3}
BR(J/\psi\to\gamma\eta(1405/1475)\to\gamma K\bar K\pi)=\nonumber
\\ =(2.8\pm0.6)\cdot10^{-3}\,,\qquad\qquad\quad
\end{eqnarray} we obtain
\begin{eqnarray}\label{Eq5-3-4} \frac{BR(J/\psi\to\gamma\eta(1405)\to
\gamma f_0(980)\pi^0\to\gamma \pi^+\pi^-\pi^0)}{BR(J/\psi\to\gamma
\eta(1405/1475)\to\gamma K\bar K\pi)}= && \nonumber \\ =(0.53\pm0.13
)\%\,.\qquad\qquad\qquad\ \ \, \end{eqnarray} This value also
implies that isospin symmetry is very strongly broken in the
$\eta(1405)$\,$\to$\,$f_0(980)\pi^0$ decay.

\begin{figure} 
\begin{center}\includegraphics[width=8cm]{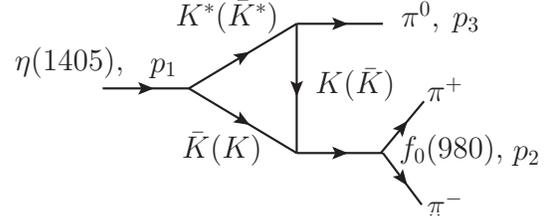}
\caption{\label{Fig5-3-1} Diagram of the $\eta(1405)\to(K^*\bar
K+\bar K^*K)\to K\bar K\pi^0\to f_0(980)\pi^0\to\pi^+\pi^-\pi^0$
decay. In the $\eta(1405)$ mass region, all intermediate particles
in the triangular loop can be on the mass shell. As a consequence, a
logarithmic singularity in the imaginary part of the
triangular-diagram amplitude emerges as soon as the $K^*$  meson is
hypothetically assumed to be stable \cite{AKS15,AK1,AK2,AK3}. The
4-momenta of corresponding particles are denoted as $p_1$, $p_2$,
and $p_3$; $p_1^2=s_1$ is the squared invariant mass of the
$\eta(1405)$ resonance or of the final $\pi^+\pi^-\pi^0$ system;
$p_2^2=s_2$ is the squared invariant mass of the $f_0(980)$
resonance or of the final $\pi^+\pi^-$ system; and
$p_3^2=m^2_{\pi^0}$.}\end{center}\end{figure}

In what follows, we will try to theoretically describe the strong
isotopic symmetry breaking in the $\eta(1405)\to
f_0(980)\pi^0\to\pi^+\pi^-\pi^0$ decay in terms of anomalous Landau
thresholds (or logarithmic triangle singularities) present in the
amplitude of the $\eta(1405)\to(K^*\bar K+\bar K^*K)\to K\bar
K\pi^0\to f_0(980)\pi^0\to\pi^+\pi^-\pi^0$ transition near the
$K\bar K$ thresholds (see Fig. 1). The authors of \cite{WLZZ12,
ALQWZ12,WWZZ13,ADO15} attempted to describe the $\eta(1405)\to
f_0(980)\pi^0\to\pi^+\pi^-\pi^0$ decay by this mechanism with the
$K^*(892)$ vector meson in the intermediate state treated as a
stable particle. Our subsequent analysis \cite{AKS15} demonstrated
that, if its finite width $\Gamma_{K^*} \approx\Gamma_{K^*\to
K\pi}\approx50$ MeV is taken into account, logarithmic singularities
in the amplitude are smoothed and the computed probability of the
$\eta(1405)\to f_0(980)\pi^0\to \pi^+\pi^-\pi^0$ decay is reduced by
a factor of 6--8 as compared to that for $\Gamma_{K^*} $\,=\,0. Also
assuming the dominance of the $\eta(1405)\to(K^* \bar K+\bar
K^*K)\to K\bar K\pi^0$ decay, we obtained in \cite{AKS15} the
estimate
\begin{eqnarray}\label{Eq5-3-5} BR(J/\psi\to\gamma\eta(1405)\to\gamma
f_0(980)\pi^0\to\gamma\pi^+\pi^-\pi^0)\approx \nonumber \\ \approx
1.12\cdot10^{-5}\,,\ \qquad\qquad\qquad\qquad \end{eqnarray} which
is in reasonably good agreement with the BESIII measurement (1).

In contrast to \cite{AKS15}, we demonstrate here in detail how the
inclusion of the nonzero $K^*$ width affects the imaginary and real
parts of the isospin-violating amplitude, effectively removing the
logarithmic singularity, and how a narrow resonance structure arises
in the $\pi^+\pi^-$ mass spectrum of the $\eta(1405)\to f_0(980)
\pi^0 \to\pi^+ \pi^- \pi^0$ decay. Apart from that, we demonstrate
for the first time that the phase of the $f_0(980)$ formation
amplitude changes abruptly by 90$^\circ$ between the $K^+K^-$ and
$K^0\bar K^0 $ thresholds.

To elucidate the impact of the nonzero $K^*$ width on the
isospin-violating transition diagrammatically shown in Fig.
\ref{Fig5-3-1}, we neglect the spin effects that significantly
complicate the intermediate calculations \cite{AKS15} but weakly
affect the final results. In other words, in what follows, the $K^*$
meson is assumed to be a spinless particle.\,\footnote{We also note
that the discussed isospin-violating effect is independent of
whether or not the triangular diagram is convergent (this equally
applies to $K\bar K$ loops in the case of the $a^0_0(980)\to(K^+K^-+
K^0\bar K^0)\to f_0(980)$ transition). This is because in the
dispersive representation of the isospin-violating amplitude, the
sum of subtraction constants for the contributions of charged and
neutral intermediate states has a natural smallness of $\sim(m_{K^0}
-m_{K^+})$ and, therefore, cannot enhance the isotopic symmetry
breaking in the narrow mass region near the $K^+K^-$ and $K^0\bar
K^0$ thresholds.}

We denote the triangle-loop amplitude (see Fig. \ref{Fig5-3-1}) as
\begin{eqnarray}\label{Eq5-3-6}
T=2\frac{g_1 g_2g_3}{16\pi}[F_+(s_1,s_2)-F_0(s_1,s_2)]\,,
\end{eqnarray}
where $g_1$, $g_2$, and $g_3$ are the coupling constants in the
three vertices assumed to be equal for the charged and neutral
channels; the amplitudes $F_+(s_1,s_2)$ and $F_0(s_1,s_2)$ describe
the contributions of the charged and neutral intermediate states,
respectively; and a factor of 2 appears because there are two such
contributions. For the discussed diagram, we assume that isotopic
symmetry breaking arises only from the mass difference between the
charged and neutral stable $K$ mesons, and set $m_{K^{*+}}=m_{K^{*0
}}=0.8955$ GeV. The amplitude $F_+\equiv F_+(s_1, s_2)$ has the form
\begin{equation}\label{Eq5-3-7}
F_+=\frac{i}{\pi^3}\int\frac{d^4k}{D_1D_2D_3}\,,
\end{equation} where
$D_1=(k^2-m^2_{K^{*+}}+i\varepsilon)$, $D_2=((p_1-k)^2-m^2_{K^-}
+i\varepsilon)$, and $D_3=((k-p_3)^2-m^2_{K^+}+i\varepsilon)$ are
the inverse propagators of the particles forming the loop. In the
region of $s_1\geq(m_{K^{*+}}+m_{K^+} )^2$ and $s_2\geq4m^2_{K^+}$,
the imaginary part of $F_+$ includes the term determined by the jump
across the $K^{*+}K^-$ cut in the $s_1$ variable and the term
determined by the jump across the $K^+K^-$ cut in the $s_2$
variable:
\begin{equation}\label{Eq5-3-8}
{\rm Im}F_+={\rm Im}F_+^{(K^{*+}K^-)}+{\rm Im}F_+^{(K^+K^-)}\,.
\end{equation} Here,
\begin{equation}\label{Eq5-3-9}{\rm
Im}F_+^{(K^{*+}K^-)}=\frac{1}{\sqrt{\Delta}}\ln\left[\frac{\alpha_+
+\sqrt{\Delta\delta_+}}{\alpha_+-\sqrt{\Delta\delta_+}}\right]\,,
\end{equation}
\begin{equation}\label{Eq5-3-10}{\rm
Im}F_+^{(K^+K^-)}=\frac{1}{\sqrt{\Delta}}\ln\left[\frac{\alpha'_+
+\sqrt{\Delta\delta'_+}}{\alpha'_+-\sqrt{\Delta\delta'_+}}\right]\,,
\end{equation} where
\begin{eqnarray}\label{Eq5-3-11}
\Delta=s^2_1-2s_1(s_2+m^2_{\pi^0})+(s_2-m^2_{\pi^0})^2,\qquad\ \ \\
\label{Eq5-3-12} \alpha_+=s^2_1-s_1(s_2+m^2_{\pi^0}
+m^2_{K^{*+}}-m^2_{K^+})+\qquad \nonumber
\\ +(s_2-m^2_{\pi^0})(m^2_{K^+}-m^2_{K^{*+}}),\qquad\qquad\  \\
\label{Eq5-3-13}
\delta_+=s^2_1-2s_1(m^2_{K^{*+}}+m^2_{K^+})+(m^2_{K^{*+}}-m^2_{K^+})^2,\\
\label{Eq5-3-14} \alpha'_+=s_2(s_2-s_1-m^2_{\pi^0}
-2m^2_{K^+}+2m^2_{K^{*+}}),\qquad \\
\label{Eq5-3-15} \delta'_+=s_2(s_2-4m^2_{K^+}).\qquad\qquad\qquad\
\end{eqnarray}
The amplitude $F_0\equiv F_0(s_1,s_2)$ is obtained by replacing the
subscript + by the subscript 0 for the functions and substituting
the masses of neutral partners for the masses of charged
intermediate particles in Eqs. (\ref{Eq5-3-7})--(\ref{Eq5-3-15}).

\begin{figure} 
\begin{center}\includegraphics[width=6.5cm]{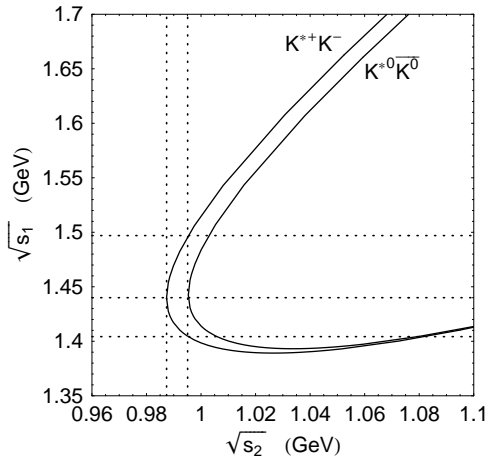}
\caption{\label{Fig5-3-2} (Solid lines) Loci of logarithmic
singularities in imaginary parts of the triangular diagram
corresponding to the contributions from the $K^{*+}K^-$ and
$K^{*0}\bar K^0$ intermediate states. Vertical dotted lines mark the
$K^+K^-$ and $K^0\bar K^0$ thresholds in the $\sqrt{s_2}$ variable
(i.e., $\sqrt{s_2}=0.987354$ GeV and $\sqrt{s_2}=0.99 5344$ GeV,
respectively). Horizontal dotted lines mark the $\sqrt{s_1}$ values
of 1.404, 1.440, and 1.497 GeV. In the $\sqrt{s_1}$ range between
1.404 and 1.497 GeV, the logarithmic singularity in the case of the
$K^{*+}K^-$ and $K^{*0}\bar K^0$ intermediate states is positioned
at $\sqrt{s_2}$ values between the $K^+K^- $ and $K^0\bar K^0$
thresholds and within 6 MeV of the $K^0\bar K^0$ threshold,
respectively. The curves touch the $K\bar K$ thresholds
approximately at $\sqrt{s_1} $\,=\,1.440 GeV.}\end{center}
\end{figure}

\begin{figure} 
\begin{center}\includegraphics[width=7cm]{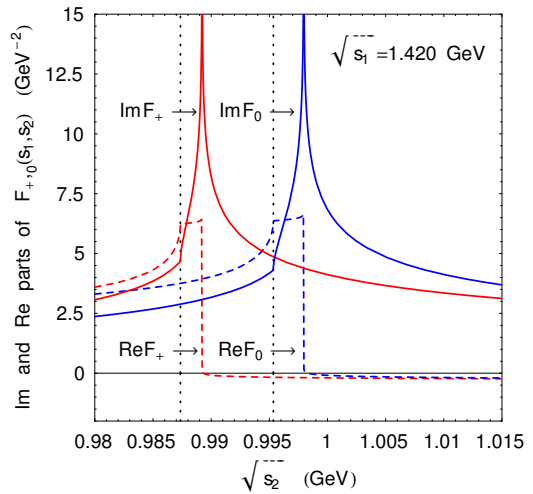}
\caption{\label{Fig5-3-3} (Color online) (Solid curves) Imaginary
and (dashed curves) real parts of the amplitudes $F_+(s_1,s_2)$ and
$F_0(s_1,s_2)$ for the charged and neutral intermediate states in
the triangular loop, respectively, computed assuming a stable $K^*$
meson in the intermediate state.}\end{center}\end{figure}

The specificity of the considered case is that all intermediate
particles in the triangular diagram in Fig. \ref{Fig5-3-1} in the
$\eta(1405)$ resonance region can be on the mass shell. This occurs
for such values of the kinematic variables $s_1$ and $s_2$ for which
\begin{eqnarray}\label{Eq5-3-16}
\alpha_{+,0}=\pm\sqrt{\Delta\delta_{+,0}}
\end{eqnarray} or, equivalently,
\begin{eqnarray}\label{Eq5-3-17}
\alpha'_{+,0}=\pm\sqrt{\Delta\delta'_{+,0}}.
\end{eqnarray}
This implies that, as soon as the $K^*$ meson is hypothetically
assumed to be stable, the amplitude of this triangular diagram has a
logarithmic singularity in its imaginary part
\cite{AKS15,AK1,AK2,AK3}. For the contributions of the intermediate
$K^{*+}K^-$ and $K^{*0}\bar K^0$ states, the loci of logarithmic
singularities in the $(\sqrt{s_2}\,,\sqrt{ s_1}\,)$ plane are shown
in Fig. \ref{Fig5-3-2}. In the $\eta(1405)$ mass region, these are
seen to be very close to the $K\bar K$ thresholds (depicted by
dotted vertical lines in this and subsequent figures). Thus, the
singularities of the $K^{*+}K^-$ and $K^{*0}\bar K^0$
intermediate-state contributions at $\sqrt{s_1}=1.420$ GeV sit at
the $\pi^+\pi^-$ invariant masses of $\sqrt{s_2}\approx0.989$ and
0.998 GeV, respectively (see Fig. \ref{Fig5-3-2}). For $\sqrt{s_1}$
in the $\eta(1405)$ mass region ($\sqrt{s_1}=1.420$ GeV for
concreteness), typical $\sqrt{s_2}$ dependences of the real and
imaginary parts of the amplitudes $F_+(s_1,s_2)$ and $F_0(s_1,s_2)$
in the $K\bar K$ threshold region are shown in Fig. \ref{Fig5-3-3}.
The dependences of $\mbox{Im}F_{+,0}(s_1,s_2)$ and
$\mbox{Re}F_{+,0}(s_1,s_2)$ have singularities and jumps,
respectively.

\begin{figure} 
\begin{center}\includegraphics[width=7cm]{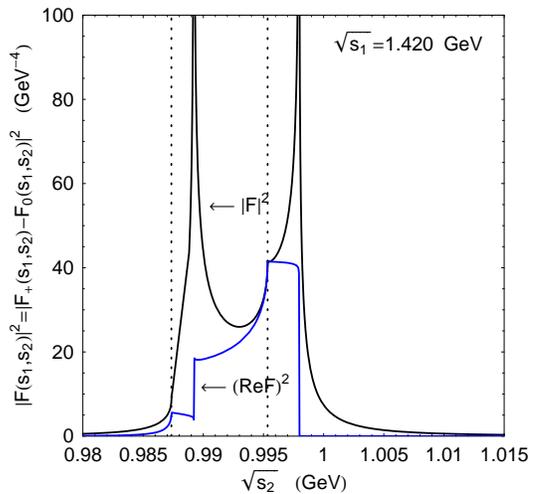}
\caption{\label{Fig5-3-4} (Color online) Squared absolute value and
squared real part of the isospin-violating triangular-loop amplitude
$F(s_1,s_2)\equiv F_+(s_1,s_2)- F_0(s_1,s_2)$ assuming a stable $K^*
$ meson in the intermediate state. Under the latter assumption,
integral contributions from the imaginary and real parts of the
amplitude are nearly equal.}\end{center}\end{figure}

Since the singularities of the $K^{*+}K^-$ and $K^{*0}\bar K^0$
contributions have different locations and do not cancel each other,
the discussed mechanism seems to induce an abrupt isotopic symmetry
breaking in the $\eta(1405)\to\pi^+\pi^-\pi^0$ decay as is
illustrated in Fig. \ref{Fig5-3-4}. However, this
singularity-dominated pattern is not realistic. This is because one
has to take into account the nonzero $K^*$ width by averaging the
amplitude over the resonant Breit-Wigner mass distribution according
to the K\"{a}ll\'{e}n-Lehmann spectral representation for the
unstable-$K^*$ propagator \cite{AK1,AK2,AK3}. This effectively
smooths the logarithmic singularities, thereby increasing the mutual
compensation of the contributions from the $(K^{*+}K^-+ K^{*-}K^+)$
and $(K^{*0}\bar K^0+\bar K^{*0}K^0)$ intermediate states. As a
result, the computed $\eta(1405)\to\pi^+\pi^-\pi^0$ width proves to
be several times less than that for $\Gamma_{K^*\to K\pi}$\,=\,0,
and the isospin-violating effect is largely restricted to
$\pi^+\pi^-$ invariant masses between the $K\bar K$ thresholds.

\begin{figure} 
\begin{center}\includegraphics[width=6.7cm]{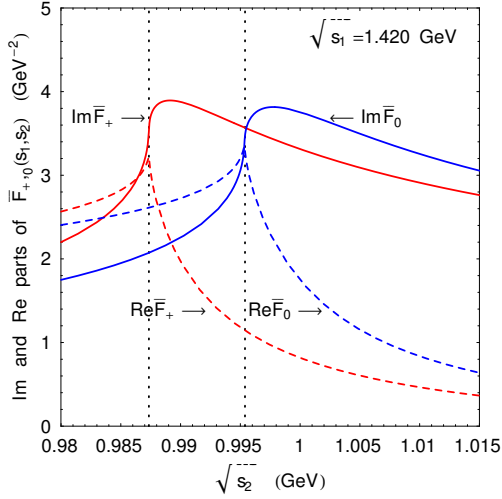}
\caption{\label{Fig5-3-5} (Color online) (Solid curves) Imaginary
and (dashed curves) real parts of the amplitudes
$\bar{F}_+(s_1,s_2)$ and $\bar{F}_0(s_1,s_2)$ for the charged and
neutral intermediate states in the triangular loop, respectively,
computed taking into account the nonzero width of the intermediate
$K^*$ meson.}\end{center}\end{figure}

\begin{figure} 
\begin{center}\includegraphics[width=9.2cm]{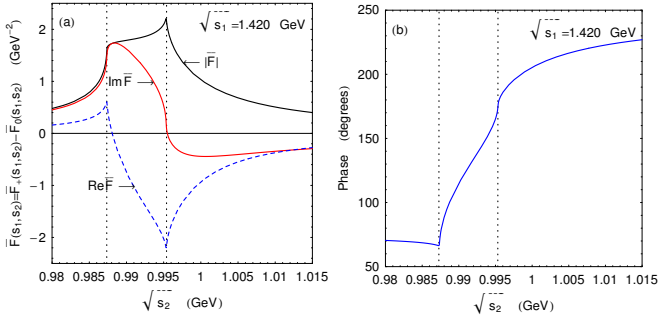}
\caption{\label{Fig5-3-6} (Color online) (a) Absolute value and the
imaginary and real parts of the triangular-loop amplitude $\bar{F}
(s_1,s_2)=\bar{F}_+(s_1,s_2)-\bar{F}_0(s_1,s_2)$ computed taking
into account the nonzero width of the intermediate $K^*$ meson. (b)
Phase of the amplitude $\bar{F}(s_1,s_2)$.}\end{center}
\end{figure}

Following this strategy, we substitute the unstable-$K^*$ propagator
in the K\"{a}ll\'{e}n-Lehmann spectral form \cite{AK1,AK2,AK3}
\begin{equation}\label{Eq5-3-18}
\frac{1}{m^2_{K^*}-k^2-im_{K^*}\Gamma_{K^*}}\to\int\limits^{\infty}_{
(m_{K}+m_\pi)^2}dm^2\frac{\rho(m^2)}{m^2-k^2-i\varepsilon}\,,
\end{equation} where $\rho(m^2)$ is approximated as
\begin{equation}\label{Eq5-3-19}
\rho(m^2)=\frac{1}{\pi}\frac{m_{K^*}\Gamma_{K^*}}
{(m^2-m^2_{K^*})^2+(m_{K^*}\Gamma_{K^*})^2}\,.
\end{equation}
Then, in the expressions for the amplitudes $F_{+,0}(s_1,s_2)$, the
$K^*$ mass squared $m^2_{K^*}$ is replaced by the variable-mass
squared $m^2$ and the amplitudes are weighted with the spectral
density $\rho(m^2)$ \cite{AK1,AK2,AK3} according to
\begin{eqnarray}\label{Eq5-3-20}
\hspace*{-0.35cm}\bar{F}_{+,0}(s_1,s_2)=\int\limits^{\infty}_{
(m_{K}+m_\pi)^2}\rho(m^2)\,F_{+,0}(s_1,s_2;m^2)\,dm^2.
\end{eqnarray}
The $\sqrt{s_2}$ dependences of the real and imaginary parts of the
weighted amplitudes $\bar{F}_+(s_1,s_2)$ and $\bar{F}_0(s_1,s_2 )$
in the $K\bar K$ threshold region are illustrated in Fig.
\ref{Fig5-3-5} for $\sqrt{s_1}=1.420$ GeV. The singularities of the
unweighted amplitudes $F_+(s_1,s_2)$ and $F_0(s_1,s_2)$ and shown in
Fig. \ref{Fig5-3-3} are seen to be practically eliminated by taking
into account the instability of the intermediate $K^*$ meson. The
absolute value, imaginary and real parts, and phase of the
isospin-violating triangle-loop amplitude $\bar{F}(s_1,s_2)\equiv
\bar{F}_+(s_1,s_2)-\bar{F}_0(s_1,s_2)$, computed taking into account
the intermediate-$K^*$ instability, are shown in Fig.
\ref{Fig5-3-6}. All characteristic irregularities of the amplitude
$\bar{F}(s_1,s_2)$ are seen to occur at the $K\bar K$ thresholds,
and its absolute value and phase behave in much the same way as
those of the $a^0_0(980)-f_0(980)$ mixing amplitude
\cite{ADS79,AS17}.

\begin{figure} [!ht]
\begin{center}\includegraphics[width=6.7cm]{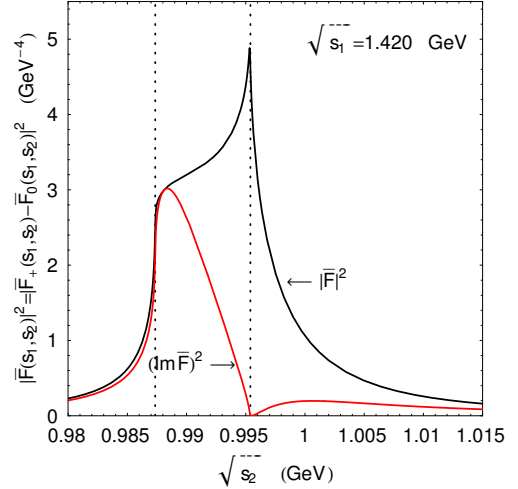}
\caption{\label{Fig5-3-7} (Color online) Squared absolute value and
squared imaginary part of the triangular-loop amplitude $\bar{F}(
s_1,s_2)=\bar{F}_+(s_1,s_2) -\bar{F}_0(s_1,s_2)$ computed taking
into account the nonzero width of the intermediate $K^*$ meson, cf.
Fig. \ref{Fig5-3-4}.}\end{center}\end{figure}

It is interesting to compare the squared absolute value of the
amplitude $\bar{F}(s_1,s_2)=\bar{F}_+(s_1,s_2)-\bar{F}_0(s_1,s_2)$
weighted with the $K^*$ spectral function to that obtained under the
assumption $\Gamma_{K^*}=0$ (cf. Figs. \ref{Fig5-3-7} and
\ref{Fig5-3-4}, respectively). Note that the areas under the
corresponding curves differ by nearly an order of magnitude, and
that this difference arises from a nonzero $K^*$ width of 50 MeV. We
also note that logarithmic triangle singularities are fully
determined by conditions (16) and (17) irrespective of particle
spins, and their modifications arising from the nonzero width are
practically unaffected by the spin effects in the $\eta(1405)\to
(K^*\bar K+\bar K^*K)\to(K^+K^-+K^0\bar K^0)\pi^0\to f_0(980)\pi^0
\to\pi^+\pi^-\pi^0$ decay. This has been explicitly demonstrated in
\cite{AKS15}.

\begin{figure} 
\begin{center}\includegraphics[width=6.7cm]{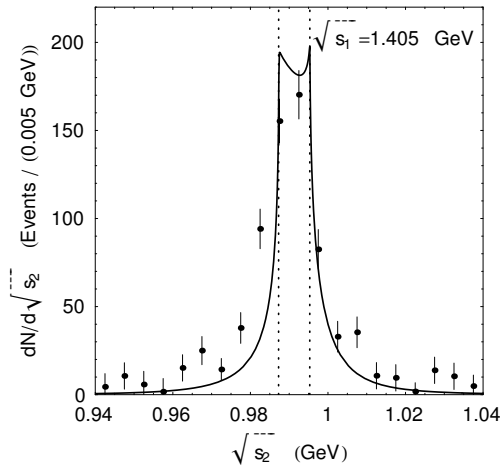}
\caption{\label{Fig5-3-8} Shape of the $\pi^+\pi^-$ mass spectrum in
the $\eta(1405)\to\pi^+\pi^-\pi^0$ decay plotted using Eq.
(\ref{Eq5-3-21}) for the contribution of the diagram in Fig.
\ref{Fig5-3-1}. Points with error bars are the first BESIII data on
this decay \cite{Ab2}.}\end{center}\end{figure}

The general pattern remains the same for all $\sqrt{s_1}$ values in
the $\eta(1405)$ mass region. Figure \ref{Fig5-3-8} shows the
general form of the $\pi^+\pi^-$ mass spectrum in the $\eta(1405)\to
\pi^+ \pi^-\pi^0$ decay obtained for the $\eta(1405)$ nominal mass,
or for $\sqrt{s_1}=1.405$ GeV, by the formula
\begin{eqnarray}\label{Eq5-3-21}
\frac{dN}{d\sqrt{s_2}}=C\sqrt{\frac{\Delta}{s_1}}\left|\bar{F}_+
(s_1,s_2)-\bar{F}_0(s_1,s_2)\right|^2\times\nonumber \\
\times\frac{s_2\Gamma_{f_0\to\pi^+\pi^-}(\sqrt{s_2})}{\pi
|D_{f_0}(\sqrt{s_2})|^2},\qquad\qquad \end{eqnarray} where $C$ is
the normalization constant and $\Gamma_{f_0\to\pi^+\pi^-}
(\sqrt{s_2})$ and $D_{f_0}(\sqrt{s_2})$ are the $\pi^+\pi^-$ partial
width and inverse propagator of the $f_0(980)$ meson, respectively
\cite{AKS16}.

In summary, let us briefly discuss expected degrees of isotopic
symmetry breaking induced by the $K^+-K^0$ mass difference in the
amplitudes of different transitions. In the processes where isotopic
symmetry breaking is determined by mass differences within meson
isospin multiplets, this degree usually amounts to
\begin{equation}\label{Eq5-3-22}
\simeq(m_{K^0}-m_{K^+})/m_{K^0}\approx1/126\,.
\end{equation}
In the processes with isotopic symmetry breaking in the region
between the $K^+K^-$ and $K^0\bar K^0$ thresholds induced by any
mechanism for the production of an S-wave $K\bar K$ pair with a
definite isospin that is free of anomalous Landau thresholds
\cite{AS17,AKS16}, such as the $a^0_0( 980)-f_0(980)$ mixing
\cite{ADS79,AS17}, the discussed degree reaches
\begin{equation}\label{Eq5-3-23}\simeq
\sqrt{2(m_{K^0}-m_{K^+})/m_{K^0}}\approx0.127\,.
\end{equation}
For isotopic symmetry breaking in the $\eta(1405)\to
f_0(980)\pi^0\to\pi^+\pi^-\pi^0$ decay amplitude arising from
logarithmic triangle singularities in the contributions from the
$(K^*\bar K+\bar K^*K)$ intermediate states in the $\sqrt{s_2}$
region between the $K^0\bar K^0$ and $K^+ K^-$ thresholds, the
discussed degree is estimated \cite{AKS15} as
\begin{equation}
\label{Eq5-3-24}\simeq\left|\ln\left|\frac{\Gamma_{K^*}/2}
{\sqrt{m^2_{K^0}-m^2_{K^+}+\Gamma^2_{K^*}/4}}\right|\right|\approx1\,.
\end{equation}
For the nonvanishing sum of the contributions of triangular diagrams
with charged and neutral intermediate states, this estimate
consistent with Fig. \ref{Fig5-3-6}a can be obtained from, e.g., Eq.
(\ref{Eq5-3-10}) upon substituting $m^2_{K^*}-im_{K^*}\Gamma_{K^*}$
for $m^2_{K^*}$ at the singularity point. In all cases of anomalous
isotopic symmetry breaking corresponding to Eqs. (\ref{Eq5-3-23})
and (\ref{Eq5-3-24}), the phase of the isotopic symmetry-violating
amplitude varies by nearly $90^\circ$ across the region between the
$K^+K^-$ and $K^0\bar K^0$ thresholds \cite{AS17,AKS15,AKS16}. \\[0.1cm]

This work was supported in part by the Russian Foundation for Basic
Research (project no. 16-02-00065) and by the Presidium of the
Russian Academy of Sciences (project no. 0314-2015-0011).


\end{document}